\documentclass[twocolumn,pra,showpacs,amsmath,amssymb]{revtex4-1}
\usepackage{graphicx}
\usepackage{bm}
\usepackage{epstopdf}
\usepackage{epsfig}
\usepackage{dsfont}
\usepackage{amssymb}
\begin{document}

\title{Characterizing scalable measures of quantum resources}
\author{Fernando Parisio}
\email[]{parisio@df.ufpe.br}
\affiliation{Departamento de
F\'{\i}sica, Universidade Federal de Pernambuco, Recife, Pernambuco
50670-901 Brazil}

\begin{abstract}

The question of how quantities, like entanglement and coherence, depend on the number of copies of a given state $\rho$ is addressed. 
This is a hard problem, often involving optimizations over Hilbert spaces of large dimensions. Here, we propose a way to circumvent the
direct evaluation of such quantities, provided that the employed measures satisfy a self-similarity property.
We say that a quantity ${\cal E}(\rho^{\otimes N})$ is {\it scalable} if it can be described as a function of the variables 
$\{{\cal E}(\rho^{\otimes i_1}),\dots,{\cal E}(\rho^{\otimes i_q}); N\}$ for $N>i_j$, while, preserving the tensor-product structure.
If analyticity is assumed, recursive relations can be derived
for the Maclaurin series of ${\cal E}(\rho^{\otimes N})$, which enable us to determine its possible functional forms (in terms of the mentioned variables). 
In particular, we find that if 
${\cal E}(\rho^{\otimes 2^n})$ depends only on ${\cal E}(\rho)$, ${\cal E}(\rho^{\otimes 2})$, and $n$, then it is completely determined by Fibonacci polynomials,
to leading order. We show that the one-shot distillable (OSD) entanglement is well described as a scalable measure for several families of states.
For a particular two-qutrit state $\varrho$, we determine the OSD entanglement for $\varrho^{\otimes 96}$ from smaller tensorings, with an accuracy 
of $97 \%$ and no extra computational effort. Finally, we show that superactivation of non-additivity may occur in this context.

\end{abstract}
\maketitle

\section{Introduction}
\label{intro}
 If, in the future, quantum resources \cite{comm0} are to be distributed to a number of users, industrial setups would be required, that is,
large scale production of copies of standard states. 
Thus, a physically and economically relevant
question concerns how the amount of resources embodied by several copies of a state relates to those of a single copy. 
A classical illustration is the advantage obtained in the assembly-line \cite{brit} production of cars. If $C(1)$ is the cost to produce a single car and
$C(N)$ is the cost to produce $N$ cars in a row, then $C(N)<N C(1)$. Therefore, in current jargon, the cost in this case is subadditive and assembly-line 
production is economically advantageous. 

In quantum mechanics, to determine some figure of merit of cost, or the entanglement, or the coherence of a large number of copies may be a prohibitive task, 
due to the exponential growth of the Hilbert space dimension as the number of copies increases (linearly).
This justifies the interest in asymptotic results ($N\rightarrow \infty$). However, in actual situations
the number of copies is always finite and, importantly, the asymptotic regime may become dominant only for an
impracticable number of copies \cite{oneshot,natcomm,ieee}. Therefore, one cannot always evade the problem
of evaluating functions of a large, but finite number of copies. In this context,
any strategy that helps to circumvent the direct evaluation of functions whose domains are 
high-dimensional Hilbert spaces is potentially useful in all fields of quantum information. 
In this article we present one such strategy, which can be applied under circumstances
to be detailed.

In the particular case of non separability, whether or not the entanglement of $N$ 
copies of a certain state coincides with $N$ times the entanglement of a single copy, 
constitutes the additivity problem \cite{comm}.
Several quantifiers have been studied under this perspective. 
The squashed entanglement \cite{squashed} and the logarithmic negativity \cite{vidal}, for instance, are additive measures, ${\cal E}(\rho^{\otimes N})=N{\cal E}(\rho)$.
For some time, this question remained open regarding the entanglement of formation \cite{formation}, which, has been 
ultimately shown to be non-additive \cite{hastings} (see \cite{horodecki} for a broader discussion).
This is the case of most of the measures, e. g., the distillable entanglement \cite{bennett,rains}, the Schmidt number \cite{horodecki}, 
the relative entropy of entanglement \cite{vedral1}, and the geometric distance \cite{geom,nonadd}, to cite a few. 

A dramatic manifestation of non-additivity is the phenomenon of superactivation, for which, given a particular quantifier ${\cal E}$, we may find 
states $\rho$, such that ${\cal E}(\rho)=0$ and ${\cal E}(\rho\otimes \rho)>0$, as is the case of the
distillable entanglement \cite{superact,watrous}. This makes it clear that the question in the first paragraph is too limited, since 
${\cal E}(\rho^{\otimes N})$ cannot possibly depend only on ${\cal E}(\rho)$, in general. In this work we consider
the question: how the amount of resources embodied by several copies of a state relates to those of {\it fewer copies}.

To illustrate this approach, suppose, in the assembly line example, we want to know the cost of producing, say, $N=30$ automobiles. 
How to determine $C(N)$ without having to actually assemble the thirty cars?
We may take an {\it ab initio} approach, by inferring the cost of buying all components in large quantities, the time gain of having assemblers assigned to specific tasks, the electric power cost 
(as a function of produced cars), etc. With such a detailed information we would be able to get $C(30)$, within a good precision. 
This complete set of inputs may not be available, however. Alternatively, one may simply produce a single car in a day and compute the total cost $C(1)$ and,
in the following day, produce two cars obtaining the total cost $C(2)$. One may try, in a self-consistent way, to determine $C(30)$ from an
extrapolation of the observed behavior, i. e., we may consider $C(30)$ as a function of the more fundamental costs: $C(30)\approx F[C(1),C(2)]$ (or $\approx F[C(1),C(2),C(3)]$, etc). 
Note that $C(1)$ and $C(2)$ are easily measurable quantities (how much one spends {\it after} producing one and two cars).
In words, knowing the cost to assemble a set of few items may give relevant information on the cost of a large number of items.

Back to quantum mechanics, the first approach would correspond to calculations, directly employing the definition of the figure of merit under scrutiny, in large Hilbert spaces. 
We will be interested in the second approach. More specifically, we may ask what are the variables that determine ${\cal E}(\rho^{\otimes N})$ and how it depends on them. 
These variables for additive measures are, of course, $\{e_1,N\}$, ${\cal E}(\rho^{\otimes N})=Ne_1$, with $e_1={\cal E}(\rho)$. 
It seems natural to investigate whether other measures are related to more complex sets of variables, say, $\{e_1, e_2, N\}$, $e_2={\cal E}(\rho^{\otimes 2})$, 
and nonlinear functional dependencies, ${\cal E}(\rho^{\otimes N})=E^{(N)}(e_1,e_2)$. In this work we investigate quantum figures of merit
such that ${\cal E}(\rho^{\otimes N})$ can be expressed as a function of 
$\{{\cal E}(\rho^{\otimes i_1}), {\cal E}(\rho^{\otimes i_2}),\dots,{\cal E}(\rho^{\otimes i_q})\}\subset\{{\cal E}(\rho^{\otimes j})\}_{j<N}$ 
and $N$. We show that this hypothesis leads to enumerative constraints which, when supplemented by ``minimal'' analyticity requirements, 
enable us to determine the functional form of the measures, at least in some range of non-vanishing resources.
Although we use entanglement as a specific example, the same idea is valid for quantifiers of coherence 
and other measures of quantumness that depend on $\rho$ and $N$.

The paper is organized as follows: In the next section we give some useful definitions and set our notations; In section $III$ we build the central concept of scalability 
and present a nontrivial illustration, namely, we show that the one-shot distillable entanglement is scalable for a family of Bell-diagonal states. 
This example works as a proof of principle, by showing that quantities which are presently under use or investigation may present scalability. 
In section $IV$ we study the large set of measures that present the elementary feature of being analytic around zero resources. With this mild restriction, we demonstrate that the possible functional dependences of ${\cal E}(\rho^{\otimes N})$ on $\{{\cal E}(\rho^{\otimes i_1}), {\cal E}(\rho^{\otimes i_2}),\dots,{\cal E}(\rho^{\otimes i_q})\}\subset\{{\cal E}(\rho^{\otimes j})\}_{j<N}$ 
are strongly constrained, through recurrence relations whose forms are presented. In this section we also test our analytical results against recent numerical computations for the OSD entanglement of a family of spherically symmetric states.  In the last section we present our final remarks and some perspectives. To improve readability, 
some technical developments and other details, which are not essential in a first reading, are given in several appendices at the end of the manuscript.
\section{Preliminary definitions and notation}

If $S$ is an arbitrary physical system (at this point it is not necessary to assume that it is quantum) and $E$ is some quantity that can be calculated from
$S$, we express this same quantity, when extended to $N$ identical (non-interacting) copies of $S$, as $E^{(N)}$. Let us denote the $q$ numbers
$(E^{(i_1)}, E^{(i_2)}, \dots, E^{(i_q)})$ by $(e_{i_1}, e_{i_2}, \dots, e_{i_q})$. 

In what follows we represent an arbitrary subset of the natural numbers as 
$\mathds{S_N}$.

{\it Definition} 1: Let $E^{(N)}$ be a function that is intended to represent a quantity associated with $N$ copies of a certain physical system $S$.
If there exists an ordered set of $q$ positive integers $\{i_1, i_2,\dots, i_q\}$ for $i_q<N$, such that
one can express $E^{(N)}$ as a function of $(e_{i_1}, e_{i_2}, \dots, e_{i_q})\equiv {\mathbf e}$,
and $N \in \mathds{S_N}$:
\begin{equation}
\label{def}
E^{(N)}(e_{i_1},e_{i_2},\dots, e_{i_q})= E^{(N)}(\mathbf{e}),
\end{equation}
we say that $E^{(N)}$ is $q$-extensible ($q$-$E$) with respect to $S$ and $\mathds{S_N}$. We assume that $q$ is the smallest integer that makes (\ref{def}) valid. 
In words if $E^{(N)}$ is $q$-$E$ its value for any large $N \in \mathds{S_N}$  is completely determined by the values $E^{(j)}$ takes for certain smaller numbers of copies. 
The total energy of $N$ non-interacting systems is $E^{(N)}=N E^{(1)}=Ne_1$, therefore, energy is 1-extensible. We will see less trivial examples in what follows.
Finally, we remark that the set $\{i_1, i_2,\dots, i_q\}$ does not coincide with
$\{1, 2,\dots, q\}$, in general. For instance, we may have $\{i_1, i_2,\dots, i_q\}=\{2, 4,\dots,2^q\}$ or $\{i_1, i_2,\dots, i_q\}=\{1, 3,\dots,2q-1\}$, etc. The referred set is fixed and,
that is, it doesn't depend on $N$ for a fixed system $S$.

Now we turn our attention to quantum systems and
consider the different ways one can group $\rho$, an arbitrary state in the Hilbert-Schmidt space ${\cal B(H)}$, to express the {\it same} tensoring
$\rho^{\otimes N}$:
\begin{eqnarray}
\nonumber
\rho^{\otimes N}&=&\rho^{\otimes N/2}\otimes\rho^{\otimes N/2}\\
\nonumber
&=&\rho^{\otimes N/4}\otimes\rho^{\otimes N/4}\otimes\rho^{\otimes N/4}\otimes\rho^{\otimes N/4}\cdots\\
\cdots&=&\rho\otimes\rho\cdots\rho\otimes\rho,
\nonumber
\end{eqnarray}
with $\rho^{\otimes J} \in {\cal B(H)}^{\otimes J}$. In these equalities, we assumed that the number of copies $N$ is a power of 2, $N=2^n$, so that
the numbers $N/2$, $N/4$, ..., $N/2^{n-1}$, $N/2^n$ are {\it always} integers, therefore, representing actual states of physical systems. 

In general, given an arbitrary positive integer $\mathfrak{a}$, one can take $N=\mathfrak{a}^n$ and $K=\mathfrak{a}^k$, with
$n,k \in \{0,1,2,\dots\}=\mathds{N}$, $k\le n$, such that
\begin{equation}
\nonumber
\rho^{\otimes N}=\sigma^{\otimes (N/K)}, \;\mbox{with}\; \sigma\equiv\rho^{\otimes K}.
\end{equation}
is well defined. In particular, we must have
\begin{equation}
\label{eq}
{\cal F}(\rho^{\otimes N})={\cal F}(\sigma^{\otimes (N/K)}),
\end{equation}
for an arbitrary function ${\cal F}$.
In addition, it will be convenient to denote the set of all integer powers of $\mathfrak{a}$  by
$$\mathds{P}_{\mathfrak{a}}=\{1,\mathfrak{a}, \mathfrak{a}^2, \dots\}$$
and use the notation $e_1={\cal E}(\rho)$, $e_2={\cal E}(\rho^{\otimes 2})$, ...,  $e_{i}={\cal E}(\rho^{\otimes i})$.

It is well known that we may have ${\cal E}(\rho^{\otimes N})=0$, even when the state $\rho^{\otimes N}$ does contain some finite amount of the considered quantity,
for instance, the negativity \cite{vidal} of bound entangled states $\rho_b$ vanishes, although $\rho_b$
is not separable.
We reserve the term ``zero-resource state'' for those states that indeed contain no resource, e. g.,
separable states for entanglement, incoherent states for coherence, and so on.

We will denote measures which are functions of $e_1$ and $N$ only, by 
\begin{equation}
\label{def0}
\mathcal{E}(\rho^{\otimes N})=E^{(N)}\left(\mathcal{E}(\rho)\right)=E^{(N)}(e_1).
\end{equation}
By definition, the condition $E^{(1)}(e_1)=e_1$ must be satisfied. For any quantifier we will assume that $\mathcal{E}(\rho^{\otimes N})=0$ 
for zero-resource states $\rho$ (but not the other way around). 
Note that, while ${\cal E}$ maps states $\rho$ in the Hilbert-Schmidt space ${\cal B(H)}$ into non-negative real numbers, 
$E^{(N)}$ takes subsets of $\mathds{R}_{+} \times \mathds{N}$ into subsets of $\mathds{R}_{+}$,

\section{Scalability}
In this section we present the concept of scalability.
Let ${\cal E}(\rho^{\otimes N})$ be a quantity such that  ${\cal E}: {\cal B(H)}^{\otimes N} \mapsto \mathds{R}_{+}$, $N \in \mathds{S_N}$.
In most of the results to be presented hereafter we will set $\mathds{S_N} =\mathds{P}_{\mathfrak{a}}$.
As we remarked, several quantifiers cannot be described via (\ref{def0}), e. g., those that allow for superactivation.
In addition, as we will see, not all $q$-$E$ functions respect the structure of the tensor product, and, thus,
are not physically acceptable. Although the tensor-product properties are built-in in first-principle quantum mechanical definitions,
if one is to take a shortcut via extensibility, these properties, Eq. (\ref{eq}) in particular, must be {\it verified}. 
In this regard, there is a simple, but non-trivial constraint that follows from the previous considerations and from 
Eq. (\ref{eq}). For the sake of clarity, we initially refer to 1-$E$ measures with $e_{i_1}=e_1$. 

{\it Proposition} 1: Let $N, K  \in \mathds{P}_{\mathfrak{a}}$, with $K < N$. Let
$\rho$ be an arbitrary quantum state and ${\cal E}$ an 1-$E$ function respecting relation (\ref{eq}), then 
\begin{equation}
\label{recur}
E^{(N)}(e_1)=E^{(N/K)}\left(E^{(K)}(e_1)\right).
\end{equation}   

{\it Proof}: Given the equality (\ref{eq}), it follows immediately that
${\cal E}(\rho^{\otimes N})={\cal E}\left(\sigma^{\otimes (N/K)}\right),$
which, in the notation introduced in (\ref{def0}), corresponds to
$E^{(N)}(e_1)={E}^{(N/K)}\left({\cal E}(\sigma)\right).$
One can use relation (\ref{def0}) again to write ${\cal E}(\sigma)={\cal E}(\rho^{\otimes K})=E^{(K)}(e_1)$ $\square$

To give a simple example of the kind of constraint the previous result imposes, consider a hypothetical $1$-$E$ function $E^{(N)}(e_1)=N(N+1)/2\,e_1$.
At first glance it seems a natural candidate as a consistent quantifier of a physical quantity related to $N$ copies of a certain system. 
Note, in particular, that  $E^{(0)}(e_1)=0$ and $E^{(1)}(e_1)=e_1$, as it should be. However, according to proposition 1, the right-hand side of Eq. (\ref{recur}),
\begin{eqnarray}
\nonumber
E^{(N/K)}\left(E^{(K)}(e_1)\right)=E^{(N/K)}\left(\frac{K(K+1)}{2}\,e_1\right)\\
\nonumber
=\frac{N}{K}\frac{(N/K+1)}{2}\frac{K(K+1)}{2}\,e_1,
\end{eqnarray}
should coincide with $E^{(N)}(e_1)=N(N+1)/2\,e_1$, which is clearly not the case. Therefore, this function is not
physically acceptable.

On the other hand, additive measures trivially satisfy relation (\ref{recur}), which, however, allows for more general dependencies. An example of a nonlinear 1-$E$ function, satisfying (\ref{recur}), is $E^{(N)}(e_1)=\lambda^{1-N}(e_1)^N$, $\lambda \in \mathds{R}_+$, as the reader cab easily check. Indeed for $\lambda=1$ this describes a multiplicative measure $E^{(N)}(e_1)=(e_1)^N$ as in the case of the pure-state entanglement measure defined in \cite{dafa2,dafa1} for $N$ even. We will have more to say about the solutions of (\ref{recur}) later. 

A consequence of proposition 1 is that the function $E^{(\mathfrak{a})}(e_1)$ completely determines $E^{(N)}(e_1)$, with $N=\mathfrak{a}^n$.
To see this, set $K=\mathfrak{a}$, $k=1$, in (\ref{recur}), $E^{(N)}(e_1)=E^{(N/\mathfrak{a})}(E^{(\mathfrak{a})}(e_1))$, which leads to $E^{(\mathfrak{a}^2)}(e_1)=E^{(\mathfrak{a})}(E^{(\mathfrak{a})}(e_1))$, and $E^{(\mathfrak{a}^3)}(e_1)=E^{(\mathfrak{a}^2)}(E^{(\mathfrak{a})}(e_1))=E^{(\mathfrak{a})}(E^{(\mathfrak{a})}(E^{(\mathfrak{a})}(e_1)))$, and so on. It is immediate, via the principle of finite induction, that 
\begin{equation}
\label{corolary0}
E^{(N)}(e_1)=\overbrace{E^{(\mathfrak{a})}\circ E^{(\mathfrak{a})}\circ  \cdots \circ E^{(\mathfrak{a})}\circ E^{(\mathfrak{a})}}^{n\; {\rm times}}(e_1),
\end{equation}
where $\circ$ denotes composition and $n=\log_{\mathfrak{a}}N$.

This means that by picking $\mathfrak{a}=2$, e. g., the behaviors of $\mathcal{E}(\rho^{\otimes 4})$, $\mathcal{E}(\rho^{\otimes 8})$, $\mathcal{E}(\rho^{\otimes 16})$, etc, are completely specified by the properties of $\mathcal{E}(\rho^{\otimes 2})$. In other words, for nonlinear 1-$E$ functions, the way $E^{(2)}(e_1)$ deviates from linearity completely determines the functions $E^{(2^n)}(e_1)$.

Now we provide the generalizations of the previous results which constrain physically consistent general $q$-extensible functions.

{\it Theorem} 1: Let $\mathfrak{a} \in \mathds{N}$, and $N, K  \in \mathds{P}_{\mathfrak{a}}$ with
$K<N$. Let $\rho \in {\cal B(H)}$ and ${\cal E}$ a $q$-$E$ function compatible with identity (\ref{eq}), i. e., 
respecting the structure of tensor products, then
\begin{equation}
\label{recur2}
E^{(N)}(\mathbf{e})=E^{(N/K)}\left(E^{(i_1K)}(\mathbf{e}), E^{(i_2K)}(\mathbf{e}), \dots, E^{(i_qK)}(\mathbf{e})\right),
\end{equation}
for $N\ge i_qK$.

{\it Proof}: The demonstration is analogous to that of proposition 1. Relation (\ref{eq}) implies ${\cal E}(\rho^{\otimes N})={\cal E}(\sigma^{\otimes N/K})$ and, since ${\cal E}$
is extensible, $E^{(N)}(\mathbf{e})=E^{(N/K)}\left({\cal E}(\sigma^{\otimes i_1}),{\cal E}(\sigma^{\otimes i_2}),{\cal E}(\sigma^{\otimes i_3}), \dots, {\cal E}(\sigma^{\otimes i_q})\right)$.
But ${\cal E}(\sigma^{\otimes j})={\cal E}(\rho^{\otimes jK})=E^{(jK)}(\mathbf{e})$ $\square$. 

There is a corollary that extends Eq. (\ref{corolary0}) to the general case. 
Since the resulting expression is cumbersome, we provide the statement and its proof in Appendix A.

{\it Definition} 2:  We say that a $q$-extensible function (with respect to $\rho$ and for all $N \in \mathds{P}_{\mathfrak{a}}$) that satisfy condition (\ref{recur2}), is a $q$-{\it scalable} ($q$-$S$) measure with respect to $\rho$ and $\mathds{P}_{\mathfrak{a}}$.

It is clear that physically consistent $q$-extensible quantum functions must be $q$-scalable. Otherwise, we might find ${\cal E}(\rho^{\otimes N})\ne{\cal E}(\sigma^{\otimes (N/K)})$, with $\sigma=\rho^{\otimes K}$, as in the example given after proposition 1.

Note that, while ${\cal E}: {\cal B(H)}^{\otimes N} \mapsto \mathds{R}_{+}$, we have
$E:\mathds{R}_{+}^{q}\times  \mathds{S_N}\mapsto  \mathds{R}_{+}$, so that, typically, 
{\it the domain of the latter has a dimension which is much lower than that of the former.}


The definition of scalability can be seen as a self-consistent generalization of additivity in two independent ways. For additive measures we simply have $E^{(N)}(e)=N e$, that is, a dependence on the single real variable $e$ {\it and} linearity. In a general $q$-scalable quantity, we may have both, several real variables $(e_{i_1}, e_{i_2}, \dots, e_{i_q})= {\mathbf e}$ and nonlinear functional dependencies. It is instructive to lift each of these constraints separately, i. e., (i) to allow for nonlinear dependences in the single variable $e$ and (ii) to consider only linear functions of the several variables 
$(e_{i_1}, e_{i_2}, \dots, e_{i_q})$. We will address these scenarios in the following sections. Another ingredient that may be considered is convexity, an essential aspect in a resource-theoretical approach. However, since we are dealing with a single state (and its tensor powers) at a time, it is not immediate how to take convexity into account in a useful way (see Appendix B for a short discussion in the context of scalable mesures).

\subsection{Example: One-shot distillation}

As a non-trivial illustration, we show that the one-shot distillable (OSD) entanglement \cite{oneshot} is an example of a $2$-$S$ function, in the regime of a large number of copies.
The OSD entanglement of a bipartite state $\rho$ is related to the maximal dimension $\kappa$ of the maximally entangled state $|\Psi\rangle=\sum |\kappa \kappa\rangle/\sqrt{\kappa}$
(within some error tolerance $\epsilon$) that can be obtained from $\rho^{\otimes N}$ via non-entangling operations, for $N$ finite (for details see \cite{ieee}). We denote this quantity 
by ${\cal E}_{OSD}^{\epsilon}(\rho^{\otimes N})$, which has been analytically determined for the family of Bell diagonal states 
$$\rho_{Bell}=p|\Psi^{+}\rangle \langle \Psi^{+}|+(1-p)|\Psi^{-}\rangle \langle \Psi^{-}|,$$
 where $|\Psi^{\pm}\rangle=(|01\rangle\pm|10\rangle)/\sqrt{2}$. The result for an arbitrary, but large number $N$ of copies reads \cite{ieee}:
\begin{eqnarray}
\nonumber
{\cal E}_{OSD}^{\epsilon}(\rho^{\otimes N}_{Bell})=N(1-h(p))\\
\nonumber
+\sqrt{Np(1-p)}\left|\log\left( \frac{1-p}{p}\right)\right|\Phi^{-1}(\epsilon)+O(\log N),
\end{eqnarray}
where $h(p)=-p\log p-(1-p) \log (1-p)$ and $\Phi^{-1}$ is the inverse of the cumulative normal distribution \cite{ieee}.

Let us consider integers $N>M>L$ sufficiently large, so that the terms $O(\log N)$ can be neglected. By setting $e_L={\cal E}_{OSD}^{\epsilon}(\rho^{\otimes L}_{Bell})\equiv e$
and $e_M={\cal E}_{OSD}^{\epsilon}(\rho^{\otimes M}_{Bell})\equiv f$, so that $L$ and $M$ are fixed, it is a simple exercise to show that 
\begin{eqnarray}
\label{OSD}
{\cal E}_{OSD}^{\epsilon}(\rho^{\otimes N}_{Bell})= \frac{(M-\sqrt{MN})\,e+(\sqrt{LN}-L)\,f}{\sqrt{LM}(\sqrt{M/N}-\sqrt{L/N})},
\end{eqnarray}
for any fixed $M$ and $L$ and arbitrary, but large $N$. Note, in particular, that for $N=L$ and $N=M$ we get the results $e$ and $f$, respectively, as it should be. 
Therefore, for a sufficiently large number of copies, the OSD entanglement can be determined as a function of the variables $e$, $f$, and $N$.
Note that, since $N>M$, the term proportional to $e={\cal E}_{OSD}^{\epsilon}(\rho^{\otimes L}_{Bell})$ gives a negative contribution
to ${\cal E}_{OSD}^{\epsilon}(\rho^{\otimes N}_{Bell})$. Since the later must be non-negative, we have $(M-\sqrt{MN})e+(\sqrt{LN}-L)f\ge 0$.

Equation (\ref{OSD}) only demonstrates that this function is 2-$E$, however, it is easy to show that it is also a $2$-$S$ measure with respect to $\rho_{Bell}$ and large integers (otherwise the definition of OSD entanglement wouldn't be consistent). 
In fact, the details of the factors multiplying the terms in $N$ and $N^{1/2}$ are irrelevant to the proof (completely analogous results hold for arbitrary pure states \cite{ieee}). 
Therefore we can state the following result.

{\it Proposition} 2: Any quantum measure that can be expressed as ${\cal E}(\rho^{\otimes N})=\mathfrak{F}N+\mathfrak{G}\sqrt{N}+O(\log N)$, where $\mathfrak{F}$ and $\mathfrak{G}$ depend on the state $\rho$ and on fixed parameters, in the limit of a large, but finite number of copies $N$, is a $2$-$S$ function up to logarithmic order.

The proposition is demonstrated in Appendix C. This constitutes a proof of principle, showing that a complex figure of merit as the OSD entanglement is a 2-scalable quantity for the whole family of Bell-diagonal states involving $|\Psi^{\pm}\rangle$, in the large-number regime.
\section{Analytic measures}

We proceed by considering functions $E^{(N)}({\bf e})$ which are analytic in the vicinity of ${\bf e}=0$.
In the  $1$-$S$ case this corresponds to the existence of a power series that converges to $E^{(N)}({ e})$
in some non-vanishing interval $[0, \epsilon_N)$, $\epsilon_N>0$. In the general $q$-$S$ case,
analyticity amounts to functions  $E^{(N)}({\bf e})$ which have a power series that converges to $E^{(N)}({\bf e})$ in 
some ball of finite radius $\epsilon_N$ (restricted to the positive hyperoctant) centred  at ${\bf e}=0$. 
Here, we will address $1$-$S$ and $2$-$S$ quantifiers, while the
treatment of general $q$-$S$ analytic functions will be deferred to a future publication.

\subsection{1-$S$ case}
Since it is clearly the most relevant case for 1-$S$ functions, we take $i_1=1$, $e_1\equiv e$ and
consider that $E^{(N)}(e)$ is analytic at $e=0$. More precisely, we will assume that the function
$E^{(N)}(e)$ has a Maclaurin series that converges in the non-vanishing interval $[0, \epsilon_N)$, $\epsilon_N>0$, 
\begin{equation}
\label{macl}
E^{(N)}(e)=d_1(N)\,e+d_2(N)\,e^2\dots=\sum_{j=1}^{\infty}d_j(N)\, e^j, 
\end{equation}
for $e \in [0, \epsilon_N]$. We defined $d_j(N)=\frac{1}{j!}\frac{d^jE^{(N)}}{de^j}\arrowvert_{e=0}, $
with  
\begin{eqnarray}
\nonumber
d_0(N)=E^{(N)}(0)&=&0 \;\; \mbox{because}\;\; e=0 \Rightarrow {\cal E}=0\\
\nonumber
  \mbox{and} \;\;d_j(1)&=&\delta_{1,j}  \;\; \mbox{because }\;\; E^{(1)}(e)=e.
 \end{eqnarray}
From the right-hand side of (\ref{recur}) we must have $E^{(N)}(e)=E^{(N/K)}\left(d_1(K)\,e+d_2(K)\,e^2\dots\right)$,
for $e \in [0,\epsilon_{K})$. By expanding $E^{(N/K)}(e)$ itself we get
\begin{eqnarray}
\nonumber
E^{(N)}(e)=d_1(N/K)\left(d_1(K)\,e+d_2(K)\,e^2\dots\right)+\\
\nonumber
d_2\left(N/K)(d_1(K)\,e+d_2(K)\,e^2\dots\right)^2+\dots\\
\nonumber
\cdots\\
d_{\ell}\left(N/K)(d_1(K)\,e+d_2(K)\,e^2\dots\right)^{\ell}+\dots
\label{macl2}
\end{eqnarray}
for $E^{(N/K)}(e) \in [0,\epsilon_{N/K})$. Comparing the terms of same order in $e$ in the last equation and in equation (\ref{macl}), we get a recursive way to determine the coefficients of the series. Specifically, for the first order coefficient we only have a single term on each side and, thus
\begin{equation}
\nonumber
 d_1(N)=d_1(N/K)d_1(K)
\end{equation}
 or, by taking the logarithm, $g_1(N)=g_1(N/K)+g_1(K)$, with $g_1(N)=\log_{\mathfrak{a}} d_1(N)$. 
 By setting $K=\mathfrak{a}$ for increasing values of $N$, it is easy to get the general solution in terms of $g_1(\mathfrak{a})$, $g_1(N)=\log_{\mathfrak{a}} N\; g_1(\mathfrak{a})$. That is
\begin{equation}
\label{fo}
 d_1(N)=\left( d_1(\mathfrak{a})\right)^n=N^{\nu},\;\; \nu\equiv \log_{\mathfrak{a}}d_1(\mathfrak{a}).
\end{equation}

Next, we compare the second order coefficients, to get the recursive relation
\begin{equation}
\nonumber
d_2(N)=\left( \frac{N}{K}\right)^{\nu}d_2(K)+K^{2\nu}d_2(N/K).
\end{equation}
The left-hand side is related to the direct series in Eq. (\ref{macl}) while the right-hand side comes from the first and second lines of Eq. (\ref{macl2}). In the above expression we already used the first order result, Eq. (\ref{fo}). The second order recursion can be easily iterated, again with $K=\mathfrak{a}$, leading to
\begin{eqnarray}
\nonumber
d_2(N)&=&\left(\frac{\left(d_1(\mathfrak{a})\right)^{n}-1}{d_1(\mathfrak{a})-1}\right)\left(d_1(\mathfrak{a})\right)^{n-1}\,d_2(\mathfrak{a}).
\end{eqnarray}
By using the first order result (\ref{fo}) in the above relation we get
\begin{equation}
\label{maclsol}
E^{(N)}(e)=N^{\nu}e+\left(\frac{N^{\nu}-1}{\mathfrak{a}^{\nu}-1}\right)\left( \frac{N}{\mathfrak{a}} \right)^{\nu}d_2(\mathfrak{a})e^2+O(e^3),
\end{equation}
for $\nu=\log_{\mathfrak{a}}d_1(\mathfrak{a})$. Therefore, we can state that {\it any} 1-$S$ measure which is analytic around $e=0$ must 
obey the above expansion. In principle, the iteration can be continued up to arbitrary order. We stress that, in order to use these relations
one must know the function $E^{(\mathfrak{a})}(e)=d_1(\mathfrak{a})e+d_2(\mathfrak{a})e^2+\cdots$, so that the coefficients are available.

Before presenting the general recurrence relation for arbitrary order we recall the definition of {\it compositions} in combinatorics. 
A composition of an integer $j$ in $\ell$ parts is an {\it ordered} sum $j=\mu_1+\mu_2+\dots+\mu_{\ell}$,
of strictly positive integers. For instance, there are 5 compositions of the integer $j=6$ in $\ell=2$ parts: 1+5, 5+1, 2+4, 4+2, and 3+3. A well-known result in enumerative combinatorics is that there are ${j-1}\choose{\ell-1}$ such compositions \cite{comb}.

{\it Theorem} 2: The general recursive relation satisfied by the Maclaurin coefficients $d_j(N)$ of 1-$S$ analytic (at $e=0$) measures is given by
\begin{equation}
\label{coef}
d_j(N)=\sum_{\ell=1}^j d_{\ell}(N/K)\sum_{i=1}^{{j-1}\choose{\ell-1}}\pi_{i}(j,\ell;K),
\end{equation}
where $\pi_{i}(j,\ell;K)=d_{\mu^i_1}(K)\,d_{\mu^i_2}(K)\dots d_{\mu^i_{\ell}}(K)$
is a product with $(\mu^i_1, \mu^i_2, \dots, \mu^i_{\ell})$ being the $i$-th composition of $j$ into $\ell$ parts.

{\it Proof}: 
The left-hand side of Eq. (\ref{coef}) comes from the series (\ref{macl}) while the right-hand side arises from (\ref{macl2}). Let us consider the terms of order $e^j$ in (\ref{macl}) and (\ref{macl2}). It is immediate that the first contributing term from (\ref{macl2}) is $d_1(N/K)d_j(K)$. Note that this term is the product of $d_{\ell=1}(N/K)$ and $\pi_{1}(j,1;K)=d_j(K)$, since there is one way to compose $j$ as the sum of a single integer ($\mu_1=j$). The last contributing term is
$d_j(N/K)\left( d_1(K)\right)^j$, which is the product of $d_{\ell=j}(N/K)$ and $\pi_{1}(j,j;K)=d_1(K)d_1(K)\dots d_1(K)=\left( d_1(K)\right)^j$, since there is only one way to compose $j$ as the sum of $j$ positive integers ($\mu_1=1, \mu_2=1,\dots, \mu_j=1$). It is clear from (\ref{macl2}) that the other contributing terms must contain products of $d$'s such that the sum of the sub-indexes equals $j$ (since these are the only terms proportional to $e^j$). The number of terms in the products is $\ell$. All possible compositions will be present, and thus, there are ${j-1}\choose{\ell-1}$ products of $d$'s for each $\ell$ \cite{comb} $\square$

As we remarked before, once we find the series $E^{(\mathfrak{a})}(e)=d_1(\mathfrak{a})e+d_2(\mathfrak{a})e^2+\cdots$,
the coefficients of the expansion of $E^{(N)}(e)$ are determined.
Note that the 1-$S$ analytical quantifiers whose regularized counterparts ${\cal E}_{reg}=\lim_{N \rightarrow \infty}{\cal E}(\rho^{\otimes N})/N$ are finite and non-zero are necessarily additive, and, thus ${\cal E}_{reg}=e$, for $e$ sufficiently small (this is not necessarily true for 2-$S$ measures).

\subsection{$2$-$S$ case}
We proceed by considering $2$-$S$ measures ${\cal E}(\rho^{\otimes N})=E^{(N)}(e,f)$, $e_1=e$ and $e_2=f$.
To alleviate the notation we set $\mathfrak{a}=2$, so that the relations that follow refer to number of copies $N=4, 8, 16$, etc.
The results derived below can be easily extended to the more general case of  $e_{\mathfrak{b}}=e$ and $e_{\mathfrak{b} \mathfrak{a}}=f$
(we will soon consider a case in which $\mathfrak{b}=6$ and $\mathfrak{a}=2$).
The measure is analytic at $(e,f)=(0,0)$ if there is a disk (restricted to $e\ge 0$ and $f\ge 0$) with finite radius around the origin such that
\begin{equation}
\nonumber
E^{(N)}(e,f)=\sum_{i,j}d_{ij}(N)e^if^j=d_{10}(N)e+d_{01}(N)f+O(2),
\end{equation}
where $O(2)$ denotes all second order terms ($\sim e^2$, $\sim f^2$, $\sim ef$).
If $\rho$ is a zero-resource state, we must have $E^{(N)}(e,f)=0$, therefore, $E^{(N)}(0,0)=d_{00}=0$ in general.
In addition, $E^{(1)}(e,f)=e$ and $E^{(2)}(e,f)=f$, implying $d_{ij}(1)=\delta_{i1}\delta_{0j}$ and 
$d_{ij}(2)=\delta_{0i}\delta_{1j}$. In the present case, theorem 1, Eq. (\ref{recur2}), reads 
$$E^{(N)}(e,f)=E^{(N/K)}\left(E^{(K)}(e,f),E^{(2K)}(e,f)\right).$$ 
With this we can state following result.

{\it Theorem} 3: If ${\cal E}$ is a 2-$S$ measure such that ${\cal E}(\rho^{\otimes 2^n})$ depends on $e={\cal E}(\rho)$,
$f={\cal E}(\rho^{\otimes 2})$ and on $n$, where ${\cal E}(\rho^{\otimes 4})=x\,e+y\,f + O(2)$, with $x$ and $y$ known, then, for
$e$ and $f$ sufficiently small, we have
\begin{equation}
\label{t3}
{\cal E}(\rho^{\otimes 2^n})=\sqrt{x^{n-1}}[\sqrt{x}F_{n-1}(\xi)\,e+F_n(\xi)\, f]+O(2),
\end{equation}
for arbitrary $n$, where $\xi=y/\sqrt{x}$, and $F_n(\xi)$ are the Fibonacci polynomials  \cite{fibonacci, sagan}. 
We note that, after inserting the explicit form of the polynomials in (\ref{t3}), only integer powers of $x$ appear
in the final expressions, thus, negative values of $x$ are possible (see Appendix D).

{\it Proof}: Using the series expansion in Eq. (\ref{recur2}) and collecting terms of first order only, we get the coupled relations
\begin{eqnarray}
\nonumber
{\cal X}(N)={\cal X}(N/K){\cal X}(K)+{\cal Y}(N/K){\cal X}(2K), \\
\nonumber
{\cal Y}(N)={\cal Y}(N/K){\cal Y}(2K)+{\cal X}(N/K){\cal Y}(K),
\end{eqnarray}
where we set ${\cal X}(N)\equiv d_{10}(N)$ and ${\cal Y}(N) \equiv d_{01}(N)$, so that $E^{(N)}(e,f)={\cal X}(N)e+{\cal Y}(N)f+O(2)$,
${\cal X}(1)=1$, ${\cal Y}(1)=0$, ${\cal X}(2)=0$, ${\cal Y}(2)=1$. Now, since $e$ and $f$ are free variables, the first 
non-trivial expansion is for $N=4$. Therefore, once we know ${\cal E}(\rho^{\otimes 4})=E^{(4)}(e,f)=\sum d_{ij}(4)e^if^j$,
then ${\cal E}(\rho^{\otimes N})$, $N>4$ is determined. For this reason we set $x={\cal X}(4)$ and $y={\cal Y}(4)$.

To derive expressions for all $N=2^n$ it suffices 
to set $k=1$, $K=\mathfrak{a}=2$. With this, the recurrence relations simplify to
\begin{eqnarray}
\nonumber
{\cal X}(N)=x{\cal Y}(N/2), \\
\nonumber
{\cal Y}(N)=y{\cal Y}(N/2)+{\cal X}(N/2).
\end{eqnarray}
By decoupling the ${\cal Y}$ relation we get the linear 
homogeneous recurrence relation with constant coefficients 
$${\cal Y}_n=y{\cal Y}_{n-1}+x{\cal Y}_{n-2},$$ 
with ${\cal Y}_j\equiv {\cal Y}(2^j)$. 
By setting ${\cal Y}_n=\sqrt{x^{n-1}}F_n$, we get the one-parameter recurrence relation
$F_n(\xi)=\xi F_{n-1}(\xi)+F_{n-2}(\xi)$,
which defines the Fibonacci polynomials \cite{fibonacci, sagan}, with $\xi=y/\sqrt{x}$
[the famous Fibonacci numbers are given by $F_n(1)$].
Therefore, by setting ${\cal E}(\rho)=e$, ${\cal E}(\rho^{\otimes 2})=f$, knowing $x$ and $y$ in 
${\cal E}(\rho^{\otimes 4})=E^{(4)}(e,f)=x\,e+y\, f$, and 
recalling that ${\cal X}_n=x{\cal Y}_{n-1}$, we get (\ref{t3})$ \square$

The explicit form of the polynomials is
\begin{eqnarray}
\nonumber
F_n(\xi)=\sum_{j=0}^{\lfloor(n-1)/2\rfloor} {{n-j-1}\choose{j}} \, \xi^{n-2j-1},
\end{eqnarray}
where $\lfloor \dots\rfloor$ denotes the floor function.  

It is not easy to find sufficient data in the literature to put these formulas to test. Recently, however, the authors of \cite{ieee} managed to numerically calculate, via linear programming, 
the OSD entanglement of up to $N=100$ copies of the symmetric state 
$$\rho_F=F\rho_d+\frac{(1-F)}{(d^2-1)}(\mathds{1}-\rho_d),$$ 
where $\rho_d$ is the maximally entangled states 
in $d$ dimensions. The parameters used in the calculations are $d=3$, $F=0.9$, and an error tolerance of $\epsilon=0.001$ \cite{ieee}. For these parameters the first non-vanishing result was obtained for six copies. Since this same figure of merit has displayed 2-scalability for Bell-diagonal and pure states, we will assume, as a working hypothesis, the same property here.

It turns out that the validity of theorem 1 can be easily extended to $N \in \{\mathfrak{b},\mathfrak{b}\mathfrak{a},\mathfrak{b} \mathfrak{a}^2, \dots\}\equiv \mathds{P}_{\mathfrak{a}}^{\mathfrak{b}}$ with $K \in \mathds{P}_{\mathfrak{a}} $ [see relation (\ref{eq})].
By setting 
\begin{equation}
\nonumber
N= 6\times 2^{n} \in \mathds{P}_{2}^{6},
\end{equation} 
using
\begin{equation}
\nonumber 
E^{(N)}({\bf e})=E^{(N/K)}(E^{(6K)}({\bf e}), E^{(12K)}({\bf e})),
\end{equation}
with ${\bf e}=(e_6,e_{12})\equiv (e,f)$, and $K=2^k \in  \mathds{P}_{2}$, we get exactly the result of Eq. (\ref{t3}), with the left-hand side replaced by ${\cal E}(\rho^{\otimes 6\times2^n})$.
We, therefore, have $e={\cal E}_{OSD}^{\epsilon}(\rho^{\otimes 6}_F)$, $f={\cal E}_{OSD}^{\epsilon}(\rho^{\otimes 12}_F)$, $x\, e+y\, f={\cal E}_{OSD}^{\epsilon}(\rho^{\otimes 24}_F)$, and 
$xy\,e+(y^2+x)\,f={\cal E}_{OSD}^{\epsilon}(\rho^{\otimes 48}_F)$. We used the numeric values for $N=6, 12, 24, 48$ in \cite{ieee} to determine $e$, $f$, $x$, and $y$ 
from these equations and predict the value of (see Appendix E):
$${\cal E}_{OSD}^{\epsilon}(\rho^{\otimes 96}_F)=x(y^2+x)e+(y^3+2xy)f= (0.7\pm 0.1)\times 96.$$
The value obtained directly from \cite{ieee} is $0.683\pm 0.001$, which corresponds to an agreement of $97\%$ with the central value predicted by formula
(\ref{t3}). This is a quite precise results since $F=0.9$ may not be exactly in the regime of weak resources, and no free parameter has been adjusted.
It would be interesting to investigate data with distinct values of $F$ and also to develop second order recurrences for 2-$S$ monotones.
Finally, we note that superactivation of non additivity may happen in $q$-$S$ functions. In the present case we may have
${\cal E}(\rho^{\otimes 2})=2{\cal E}(\rho)$, that is, $f=2e$, however, with ${\cal E}(\rho^{\otimes 4})=(x+2y)\,e \ne 4{\cal E}(\rho)$
whenever $x\ne 4-2y$.

\section{Closing remarks}

Any strategy that helps to circumvent the direct evaluation of functions whose domains are high-dimensional Hilbert spaces may be useful in several fields of quantum information. In this work we have introduced the concept of scalability for any physical figure of merit which is solely determined by the $N$-fold tensor product of a quantum state $\rho$. Although we referred to entanglement, the presented results are equally valid for coherence measures \cite{coh,coh2,coh3,coh4,coh5,coh6,coherence}, for example, with minor adaptations (e. g., replacing ``vanishing on separable states'' with ``vanishing on incoherent states''). This approach enabled us to employ elementary tools from analysis and combinatorics to the study of a broad class of quantum functions. The introduced concepts and the consequent results may be naturally seen as a generalization of the utterly restrictive idea of additivity.

A more extensive numeric investigation of the OSD entanglement and a study of scalability properties of coherence measures ere in progress.
A potentially interesting approach is to consider the description of physical quantities via scalable functions as an approximative method. 
Given that a certain quantity is {\it not} scalable, how well can we approximate it via a scalable expression? 
Another point that will be investigated in the future is, can we define easily computable measures by using scalability as
an ingredient from the outset?

It is worth mentioning that the same ideas may also be extended to quantities that depend on information not contained in $\rho^{\otimes N}$. Several measures of Bell nonlocality \cite{KL, bell, fonseca} and steering \cite{renato}, e. g., refer to the state plus the Bell scenario, which can fit into the presented formalism provided that the number of observables per party remains fixed as the number of copies grow. For direct quantifications of quantum behaviors \cite{chaves} the extension is not immediate.

\begin{acknowledgments}
The author thanks Marco Tomamichel for his kind assistance in the clarification of some points of reference \cite{ieee}. This work received financial support from the Brazilian agencies Coordena\c{c}\~ao de Aperfei\c{c}oamento de Pessoal de N\'{\i}vel Superior (CAPES), Funda\c{c}\~ao de Amparo \`a Ci\^encia e Tecnologia do Estado de Pernambuco (FACEPE), and Conselho Nacional de Desenvolvimento Cient\'{\i}fico  e Tecnol\'ogico through its program CNPq INCT-IQ (Grant 465469/2014-0).
\end{acknowledgments}

\appendix
\begin{widetext}
\section{Corollary 1}

{\it Corollary} 1: The function $E^{(N)}({\mathbf e})$ is completely determined by  $E^{(i_1\mathfrak{a})}({\mathbf e})$, $E^{(i_2\mathfrak{a})}({\mathbf e})$, ..., $E^{(i_q\mathfrak{a})}({\mathbf e})$:
\begin{equation}
E^{(N)}(\mathbf{e})\hspace{-0.1cm}=\hspace{-0.1cm}E^{(i_q\mathfrak{a})}\hspace{-0.1cm} \left(\overbrace{\hspace{-0.1cm}E^{(i_1\mathfrak{a})}\left(\hspace{-0.1cm}\right.E^{(i_1\mathfrak{a})}\dots (E^{(i_1\mathfrak{a})}}^{(n-2)\; {\rm times}}(E^{(i_1\mathfrak{a})}({\mathbf e}),\right.\dots, E^{(i_q\mathfrak{a})}({\mathbf e})), \dots, \left. \overbrace{E^{(i_q\mathfrak{a})}\left(\hspace{-0.1cm} \right.E^{(i_q\mathfrak{a})}\dots (E^{(i_q\mathfrak{a})}}^{(n-2)\; {\rm times}}(E^{(i_1\mathfrak{a})}({\mathbf e}), \dots, E^{(i_q\mathfrak{a})}({\mathbf e}))\dots)\hspace{-0.1cm}\right)\hspace{-0.1cm}.
\label{corollary1}
\end{equation}
{\it Proof}: By setting $K=\mathfrak{a}$ and for
$N=i_q\mathfrak{a}^2, i_q\mathfrak{a}^3, \dots$ we get $E^{(i_q\mathfrak{a}^2)}=E^{(i_q\mathfrak{a})}(E^{(i_1\mathfrak{a})}(\mathbf{e}), E^{(i_2\mathfrak{a})}(\mathbf{e}), \dots, E^{(i_q\mathfrak{a})}(\mathbf{e}))$
and 
\begin{eqnarray}
\nonumber
E^{(i_q\mathfrak{a}^3)}&=&E^{(i_q\mathfrak{a}^2)}\left(E^{(i_1\mathfrak{a})}(\mathbf{e}), \dots, E^{(i_q\mathfrak{a})}(\mathbf{e})\right)\\
\nonumber
&=&E^{(i_q\mathfrak{a})}\left( E^{(i_1\mathfrak{a})}(E^{(i_1\mathfrak{a})}(\mathbf{e}), \dots, E^{(i_q\mathfrak{a})}(\mathbf{e})),  \dots , E^{(i_q\mathfrak{a})}(E^{(i_1\mathfrak{a})}(\mathbf{e}), \dots, E^{(i_q\mathfrak{a})}(\mathbf{e})) \right).
\end{eqnarray} 
Using theorem 1 again we get
\begin{eqnarray}
\nonumber
E^{(i_q\mathfrak{a}^4)}&=&E^{(i_q\mathfrak{a}^3)}\left(E^{(i_1\mathfrak{a})}(\mathbf{e}), \dots, E^{(i_q\mathfrak{a})}(\mathbf{e})\right)\\
\nonumber
&=&E^{(i_q\mathfrak{a})}\left( E^{(i_1\mathfrak{a})}(E^{(i_1\mathfrak{a})}(E^{(i_1\mathfrak{a})}(\mathbf{e}), \dots, E^{(i_q\mathfrak{a})}(\mathbf{e})),  \dots , E^{(i_q\mathfrak{a})}(E^{(i_q\mathfrak{a})}(E^{(i_1\mathfrak{a})}(\mathbf{e}), \dots, E^{(i_q\mathfrak{a})}(\mathbf{e})) \right),
\end{eqnarray} 
and so on.
It is immediate that by assuming the validity of eq. (6) of the main text for $N=i_q\mathfrak{a}^n$ it will be valid for $N=i_q\mathfrak{a}^{n+1}$. Since corollary 1 is trivially valid for $N=i_q\mathfrak{a}$, due to the principle of finite induction, we get (\ref{corollary1}) $\square$

\section{Convexity}

Convexity is a pervasive geometric feature of the quantum formalism \cite{bengtsson}. Physically, it may represent the constraint that one cannot enlarge the amount of resources by mixing: ${\cal E}(q\varrho+(1-q)\sigma)\le q{\cal E}(\varrho)+(1-q){\cal E}(\sigma)$, $0\le q\le 1$. Since, in this work, we are always referring to a single state and its tensor powers, it is not immediate how to take convexity into account. A first step is to consider the convex combination with $\varrho=\rho^{\otimes N}$ and $\sigma=\rho_{wn}^{\otimes N-K}\otimes \rho^{\otimes K}$, where $\rho_{wn}=\mathds{1}/d$ is the completely unpolarized state (white noise), $D={\rm dim}({\cal H})$. In this case, convexity reads:
\begin{eqnarray}
\nonumber
{\cal E}(q\varrho+(1-q)\sigma)&\le& q{\cal E}(\rho^{\otimes N})+(1-q){\cal E}(\rho_{wn}^{\otimes N-K}\otimes \rho^{\otimes K})\\
&=&qE^{(N)}(\mathbf{e})+(1-q)E^{(K)}(\mathbf{e}),
\end{eqnarray}
where, in the second line, we are restricted to quantifiers for which white noise represents no resource. By using the explicit form of theorem 1, convexity assumes the form
\begin{eqnarray}
\nonumber
{\cal E}(q\varrho+(1-q)\sigma)&\le& qE^{(N/K)}\left(E^{(i_1K)}(\mathbf{e}), E^{(i_2K)}(\mathbf{e}), \dots, E^{(i_qK)}(\mathbf{e})\right)+(1-q)E^{(K)}(\mathbf{e}).
\end{eqnarray}
If we consider the simplest case of $q=1$ and $i_1=1$ we get
\begin{eqnarray}
\nonumber
{\cal E}(q\varrho+(1-q)\sigma) \le qE^{(N/K)}(E^{(K)}(e))+(1-q)E^{(K)}(e).
\end{eqnarray}
%

\section{Proof of proposition 2}

\vspace{0.5cm}

From the definitions in the main text, we have 
$$\mathfrak{F}=\frac{\sqrt{L}f-\sqrt{M}e}{M\sqrt{L}-L\sqrt{M}} \;\;\mbox{and} \;\; \mathfrak{G}=\frac{Lf-Me}{L\sqrt{M}-M\sqrt{L}}.$$ 
With these relations one can explicitly write Eq. (7) in the main text as
\begin{equation}
\label{OSDA}
E^{(N)}(e,f)=\frac{\sqrt{N}}{\sqrt{M}-\sqrt{L}}\left[ \left(\frac{\sqrt{M}-\sqrt{N}}{\sqrt{L}} \right)e+ \left(\frac{\sqrt{N}-\sqrt{L}}{\sqrt{M}} \right)f\right].
\end{equation}
It is a simple exercise to show that the above relation is valid up to logarithmic order for any resource function which can be written as ${\cal E}(\rho^{\otimes N})=\mathfrak{F}N+\mathfrak{G}\sqrt{N}+O(\log N)$. The proof of proposition 2 consists in showing that the above function satisfy theorem 1, Eq. (6) in the main text, that is
\begin{eqnarray}
\nonumber
E^{(N)}(e,f)&=&E^{(N/K)}\left(E^{(KL)}(e,f),E^{(KM)}(e,f)\right)\\
\nonumber
&=&\frac{\sqrt{N/K}}{\sqrt{M}-\sqrt{L}}\left( \left(\frac{\sqrt{M}-\sqrt{N/K}}{\sqrt{L}} \right)E^{(KL)}(e,f)+ \left(\frac{\sqrt{N/K}-\sqrt{L}}{\sqrt{M}} \right)E^{(KM)}(e,f)\right)\\
\nonumber
&=&\frac{1}{(\sqrt{M}-\sqrt{L})^2}\left\{ \sqrt{NL}\left(\frac{\sqrt{M}-\sqrt{N/K}}{\sqrt{L}} \right)\left[ \left(\frac{\sqrt{M}-\sqrt{KL}}{\sqrt{L}} \right)e+ \left(\frac{\sqrt{KL}-\sqrt{L}}{\sqrt{M}} \right)f\right]\right.\\
\nonumber
&+&\left. \sqrt{KM}\left(\frac{\sqrt{N/K}-\sqrt{L}}{\sqrt{M}} \right)\left[ \left(\frac{\sqrt{M}-\sqrt{KM}}{\sqrt{L}} \right)e+ \left(\frac{\sqrt{KM}-\sqrt{L}}{\sqrt{M}} \right)f\right]\right\}.
\end{eqnarray}
With some further algebraic manipulations one shows that all terms containing $K$ cancel out and, in addition, that the remaining terms exactly coincide with Eq. (\ref{OSDA}) above, which finishes the proof.

\section{Some explicit formulas for equation (12)}
\vspace{0.5cm}

Let us consider a 2-$S$ measure, for which one can express ${\cal E}(\rho^{\otimes \mathfrak{b}})=e$, ${\cal E}(\rho^{\otimes 2\mathfrak{b}})=f$, and ${\cal E}(\rho^{\otimes 4\mathfrak{b} })=x e+ y f +O(2)$, for a constant integer $\mathfrak{b}$. This is a slightly more general case for which theorem 1 can be extended trivially. Application of theorem 3, Eq. (13) in the main text, leads to
$${\cal E}(\rho^{\otimes 8\mathfrak{b}})=xye+(y^2+x)f +O(2),$$
$${\cal E}(\rho^{\otimes 16\mathfrak{b}})=x(y^2+x)e+(y^3+2xy)f+O(2),$$
$${\cal E}(\rho^{\otimes 32 \mathfrak{b}})=x(y^3+2xy)e+(y^4+3xy^2+x^2)f+O(2),$$
$${\cal E}(\rho^{\otimes 64 \mathfrak{b}})=x(y^4+3xy^2+x^2)e+(y^5+4xy^3+3x^2y)f+O(2),\; etc,$$
where $O(2)$ denotes terms proportional to $e^2$, $ef$, and $f^2$.

\section{Numeric parameters used to estimate ${\cal E}_{OSD}^{\epsilon}(\rho^{\otimes 96}_F)$}
\vspace{0.5cm}

We used a software to carefully extract the data from a zoomed copy of figure 2 of reference [3] of the main text, for $N=6,7, \dots, 99,100$. 
In particular the OSD entanglement per copy for the four points mentioned in the main text is $$\frac{{\cal E}_{OSD}^{\epsilon}(\rho^{\otimes 6}_F)}{6}= 0.167\pm0.001, \;\;\frac{{\cal E}_{OSD}^{\epsilon}(\rho^{\otimes 12}_F)}{12}= 0.308 \pm 0.001,\;\;
\frac{{\cal E}_{OSD}^{\epsilon}(\rho^{\otimes 24}_F)}{24}= 0.439\pm 0.001,\;\; \frac{{\cal E}_{OSD}^{\epsilon}(\rho^{\otimes 48}_F)}{48}= 0.573\pm 0.001.$$
From the relations 
\begin{equation}
e={\cal E}_{OSD}^{\epsilon}(\rho^{\otimes 6}_F),\; f={\cal E}_{OSD}^{\epsilon}(\rho^{\otimes 12}_F),\; e x+ f y={\cal E}_{OSD}^{\epsilon}(\rho^{\otimes 24}_F), \; e xy+f(x+y^2)={\cal E}_{OSD}^{\epsilon}(\rho^{\otimes 48}_F),
\end{equation}
we get
\begin{equation}
x=\frac{{\cal E}_{OSD}^{\epsilon}(\rho^{\otimes 24}_F)-f y}{e}, \; \mbox{with}\; y=\frac{e {\cal E}_{OSD}^{\epsilon}(\rho^{\otimes 48}_F)-f{\cal E}_{OSD}^{\epsilon}(\rho^{\otimes 24}_F)}{e {\cal E}_{OSD}^{\epsilon}(\rho^{\otimes 24}_F)-f^2}.
\end{equation}
Using standard uncertainty propagation formulas we obtain $x= -3.1\pm0.3$, $y= 3.71\pm0.08$. For $N=96$ we get
$$\frac{{\cal E}_{OSD}^{\epsilon}(\rho^{\otimes 96}_F)}{96}=\frac{x(y^2+x)e+(y^3+2xy)f}{96} =0.7\pm0.1\;\; \;\;{\rm(from\; theorem \;3)}.$$
The numeric value obtained from reference \cite{ieee} is 
$$\frac{{\cal E}_{OSD}^{\epsilon}(\rho^{\otimes 96}_F)}{96}\approx 0.683\pm0.001.$$
\end{widetext}


\begin{thebibliography}{99}
\bibitem{comm0} The term ``resource'' is employed in the general sense, since the
results to be derived do not rely on all the requirements for a quantity to be 
a resource (in the resource-theoretic sense).
\bibitem{comm} A stronger notion of additivity is ${\cal E}(\rho\otimes\sigma)={\cal E}(\rho)+{\cal E}(\sigma)$, for
all pairs of states $\rho$ and $\sigma$. In this manuscript whenever we refer to additivity we mean the weaker
condition.
\end{thebibliography}
\end{document}